\documentclass[conference,10pt]{IEEEtran}

\IEEEoverridecommandlockouts

\usepackage{cite}
\usepackage{amsmath,amssymb,amsfonts}
\usepackage{bm}
\usepackage{graphicx}
\usepackage{algorithm}
\usepackage{algorithmic}
\usepackage{array}
\usepackage{booktabs}
\usepackage{xcolor}
\usepackage{geometry}
\usepackage{soul}

\def\BibTeX{{\rm B\kern-.05em{\sc i\kern-.025em b}\kern-.08em
    T\kern-.1667em\lower.7ex\hbox{E}\kern-.125emX}}

\newcommand{\CN}{\mathcal{CN}}
\newcommand{\diag}{\operatorname{diag}}

\newcommand{\Real}{\operatorname{Re}}

\begin{document}
	
	\title{\huge Secrecy Rate Maximization for UAV-Mounted\\
		Six-Dimensional Movable IRS-Assisted ISAC Systems}
	
	\author{
		\IEEEauthorblockN{Chengye Hong$^{1,*}$, Botang Shi$^{1,*}$, Rongkun Zhu$^{1}$, Yuhan Wang$^{2}$, Chenyiming Jiang$^{1}$, and Lei Xie$^{1,\dagger}$}
	\textit{$^{1}$School of Cyber Science and Engineering, Southeast University, Nanjing 210096, China}\\
	\textit{$^{2}$Chien-Shiung Wu College, Southeast University, Nanjing 210096, China}\\
    \thanks{ This work was supported in part by the National Natural Science Foundation of China (No. 62501143), in part by the Basic Research Program of Jiangsu (No. BK20251332), and in part by Undergraduate Training Programs for Innovation of Jiangsu Province (No. S202610286144). $^{*}$Chengye Hong and Botang Shi contributed equally to this work. $^{\dagger}$Corresponding author: Lei Xie (email: leixie@seu.edu.cn).}
   
	}
	
	\maketitle
	
	\begin{abstract}
		Integrated sensing and communication (ISAC) is a key enabling technology for 6G wireless networks, but its broadcast nature raises a physical-layer security concern when the sensing target can act as a potential eavesdropper. Although intelligent reflecting surfaces (IRSs) can enhance wireless propagation and improve secrecy, existing secure IRS-assisted ISAC designs are mostly limited to fixed deployments and passive phase control, which offer limited spatial adaptability in line-of-sight-dominated low-altitude scenarios. To address this limitation, we investigate an unmanned aerial vehicle (UAV)-mounted six-dimensional movable IRS-assisted secure ISAC system, where the IRS location, orientation, and reflection coefficients are jointly optimized with the BS beamformer to maximize the secrecy rate under communication quality-of-service (QoS), power, unit-modulus, and visibility constraints. The resulting problem is highly non-convex due to the coupled active/passive beamforming variables and the location-and-orientation-dependent (pose-dependent) channel responses. To solve it efficiently, we develop a three-block alternating optimization (AO) framework, in which the active beamformer, IRS pose, and passive reflection vector are updated via linearized ADMM, warm-started particle swarm optimization, and Riemannian gradient descent, respectively. Simulation results show that the proposed design significantly outperforms fixed-location and orientation-only baselines, highlighting the importance of joint translation, rotation, and phase control for secure ISAC.
	\end{abstract}
	
	
	\section{Introduction}
	Integrated sensing and communication (ISAC) has emerged as one of the foundational paradigms for sixth-generation (6G) wireless networks, since it enables sensing and communication functionalities to share spectrum, hardware, and signal processing resources in a unified framework~\cite{liu2020jointradarcomm, xie2023collaborativesensing}. 
    However, the reuse of a common waveform for both sensing and communication also introduces a fundamental physical-layer security concern \cite{xie2026secure}. In particular, the object being sensed may simultaneously act as a potential eavesdropper, and thus confidential communication intended for legitimate users can be exposed through the sensing process itself. Therefore, achieving security-aware ISAC is no longer an optional enhancement, but a prerequisite for trustworthy 6G systems.

    Among the technologies developed to improve wireless propagation environments, intelligent reflecting surface (IRS) has attracted considerable attention because of its low-cost, energy-efficient, and programmable phase-control capability~\cite{wu2020towardsris}. In ISAC systems, IRS can enhance the desired communication link while suppressing the leakage toward unintended directions, thereby offering an effective means to improve both sensing and secrecy performance~\cite{ wu2025irsisacsurvey}. Nevertheless, most existing IRS-assisted secure transmission schemes rely on fixed deployment and passive phase tuning only \cite{ pang2022irssecureuav}. Such designs are particularly limited in low-altitude line-of-sight (LoS)-dominant scenarios, where the base station (BS), legitimate user, and sensing target/eavesdropper often lie in closely aligned angular directions. In these environments, fixed IRSs may not provide sufficient spatial degree of freedom to simultaneously strengthen the legitimate link and weaken the eavesdropping link. 

To overcome the inherent rigidity of conventional fixed IRS deployments, movable antenna technologies have recently been proposed as a promising new direction for wireless system design. In particular, a UAV-mounted six-dimensional movable IRS (6D movable IRS) can jointly adjust its three-dimensional position and three-dimensional orientation, thus reconfiguring not only the path loss but also the incident/reflection angles and the effective aperture projection of the cascaded channels~\cite{zhou2025rotatableirs6dma, wu2025mairs,wang2025uav}. Compared with phase-only control, such six-dimensional geometric flexibility provides a new layer of macroscopic channel control, which is especially valuable in LoS-dominated low-altitude environments where subtle changes in location and pose can lead to significant variations in channel geometry.
However, most of the current literature focuses on sum-rate maximization, coverage extension, or beam alignment, while secrecy-aware designs remain relatively underexplored. In particular, how to exploit six-dimensional mobility to improve secrecy in ISAC systems is still an open problem.

    Motivated by these observations, we propose a UAV-mounted 6D movable IRS-assisted secure ISAC system, where the BS transmits a unified orthogonal frequency-division multiplexing (OFDM) signal to serve a legitimate user while simultaneously sensing a target that may act as a potential eavesdropper. By jointly adjusting the IRS three-dimensional location, three-dimensional orientation, and element-wise reflection coefficients, the proposed architecture dynamically reshapes the cascaded propagation channels and creates favorable spatial discrimination between the legitimate communication link and the leakage link toward the target. To exploit these geometric and electromagnetic degrees of freedom, we formulate a secrecy-rate maximization problem under transmit-power, communication quality-of-service (QoS), unit-modulus reflection,  and physical visibility constraints. To tackle the resulting highly non-convex problem, we develop a three-block alternating optimization framework: the BS active beamformer is updated via a linearized alternating direction method of multipliers (ADMM) \cite{boyd2011distributed}, the six-dimensional IRS pose is searched using warm-started particle swarm optimization (PSO) \cite{kennedy1995particle}, and the passive reflection vector is optimized over the complex-circle manifold by Riemannian gradient descent. Simulation results demonstrate that the proposed scheme achieves stable convergence and significantly improves the secrecy rate compared with fixed-location and orientation-optimized IRS baselines, thereby confirming the critical role of joint displacement, rotation, and phase control in secure low-altitude ISAC systems.
	
	
	\section{System Model}
	
	\subsection{Network Architecture and Coordinate Systems}
	
	We consider a downlink secure ISAC system assisted by a UAV-mounted IRS. The BS is equipped with $N_t$ transmit antennas and $N_r$ receive antennas. A single-antenna legitimate user, denoted by UE, receives confidential information from the BS. Meanwhile, a sensing target, denoted by Eve, may act as a potential eavesdropper. The UAV carries a passive IRS with $N=N_xN_z$ reflecting elements arranged as a uniform planar array (UPA).
	
	A global Cartesian coordinate system (GCCS) is established with the BS array center as the origin. The positions of the BS, UE, Eve, and IRS center are denoted by $\mathbf p_B = [0,0,0]^T$, $\mathbf p_U = [p_{U,x},p_{U,y},p_{U,z}]^T$, $\mathbf p_E = [p_{E,x},p_{E,y},p_{E,z}]^T$, and $\mathbf p_R = [p_{R,x},p_{R,y},p_{R,z}]^T$, respectively.

A local Cartesian coordinate system (LCCS) is attached to the IRS center. The IRS \textit{orientation} is represented by the Euler-angle vector $\gamma$, while the IRS \textit{pose} is defined as the joint 6D vector consisting of its 3D position $\mathbf{p}_{\mathrm{R}}$ and orientation $\gamma$, which is characterized by
\begin{equation}
g \triangleq [\mathbf{p}_{\mathrm{R}}^T, \gamma^T]^T,
\end{equation}
where $\gamma = [\gamma_x, \gamma_y, \gamma_z]^T$ denotes the Euler-angle vector representing roll, pitch, and yaw. Let $\mathbf{Q}(\gamma) \in \mathrm{SO}(3)$ denote the rotation matrix from the LCCS to the GCCS.
    Then, for any node $X\in\{B,U,E\}$, its local direction observed from the IRS is determined by
	\begin{equation}
		\mathbf u_{R,X}^{\rm L}(\mathbf g)
		=
		\frac{\mathbf Q^T(\boldsymbol\gamma)(\mathbf p_X-\mathbf p_R)}
		{\|\mathbf p_X-\mathbf p_R\|_2}.
		\label{eq:local_direction}
	\end{equation}
	This transformation is essential because the IRS array response and aperture projection depend on the incident and reflection angles in the local IRS coordinate system.
	
	The $n$-th IRS element has local coordinate $\bar{\mathbf r}_n\in\mathbb R^3$. For a UPA lying on the $x'z'$-plane, $\bar{\mathbf r}_n$ can be written as $\bar{\mathbf{r}}_{n} = \left[ \left(m-\frac{N_x+1}{2}\right)d_x, 0, \left(\ell-\frac{N_z+1}{2}\right)d_z \right]^T$,
	where  $n=(m-1)N_z + \ell$, $m=1,\ldots,N_x$, $\ell=1,\ldots,N_z$, and $d_x,d_z$ are the element spacings.

	Let $\lambda$ denote the carrier wavelength. The IRS steering vector toward node $X$ is modeled as
	\begin{equation}
		\mathbf a_R(\mathbf u_{R,X}^{\rm L})
		=
		\frac{1}{\sqrt N}
		\left[
		e^{-\jmath \frac{2\pi}{\lambda} \bar{\mathbf r}_1^T\mathbf u_{R,X}^{\rm L}},
		\ldots,
		e^{-\jmath \frac{2\pi}{\lambda} \bar{\mathbf r}_N^T\mathbf u_{R,X}^{\rm L}}
		\right]^T.
		\label{eq:irs_steering}
	\end{equation}
	Similarly, the BS transmit steering vector toward node $X$ is denoted by $\mathbf a_B(\mathbf u_{B,X})\in\mathbb C^{N_t\times 1}$, where $
		\mathbf u_{B,X}=\frac{\mathbf p_X-\mathbf p_B}{\|\mathbf p_X-\mathbf p_B\|_2}.$
	
	\subsection{Communication Channel}
	
	Since the UAV-mounted IRS operates in a low-altitude LoS-dominant environment, all links are modeled as LoS channels. For nodes $I,J\in\{B,R,U,E\}$, define $d_{I,J}=\|\mathbf p_J-\mathbf p_I\|_2.$ 
	The corresponding complex path gain is
	\begin{equation}
		\alpha_{I,J}
		=
		\sqrt{\beta_0 d_{I,J}^{-\eta_{I,J}}}
		e^{-\jmath 2\pi d_{I,J}/\lambda},
		\label{eq:path_gain}
	\end{equation}
	where $\beta_0$ is the reference path loss at $1$ meter and $\eta_{I,J}$ is the path-loss exponent.
	
	The direct BS--UE and BS--Eve channels are expressed as
		$\mathbf h_{B,U}^H = \alpha_{B,U}\mathbf a_B^H(\mathbf u_{B,U})$, and $
		\mathbf h_{B,E}^H = \alpha_{B,E}\mathbf a_B^H(\mathbf u_{B,E})
	$, respectively.
	The BS--IRS channel is given by
	\begin{equation}
		\mathbf G_{B,R}(\mathbf g)
		=
		\alpha_{B,R}
		\mathbf a_R(\mathbf u_{R,B}^{\rm L})
		\mathbf a_B^H(\mathbf u_{B,R})
		\in\mathbb C^{N\times N_t}.
		\label{eq:BS_IRS_channel}
	\end{equation}
	The IRS-$X$ channel, where $X\in\{U,E\}$, is given by
	\begin{equation}
		\mathbf h_{R,X}^H(\mathbf g)
		=
		\alpha_{R,X}
		\mathbf a_R^H(\mathbf u_{R,X}^{\rm L}).
		\label{eq:IRS_X_channel}
	\end{equation}
	
	Let the IRS reflection matrix be
	\begin{equation}
		\boldsymbol\Theta
		=
		\diag(\mathbf v)
		=
		\diag(e^{\jmath\theta_1},\ldots,e^{\jmath\theta_N}),
		\label{eq:theta}
	\end{equation}
	where $\mathbf v=[e^{\jmath\theta_1},\ldots,e^{\jmath\theta_N}]^T$ and $|v_n|=1$.
	
	Due to the finite planar aperture of the IRS, the effective reflection gain depends on the incident and reflection angles. Let $\mathbf n(\boldsymbol\gamma)=\mathbf Q(\boldsymbol\gamma)\mathbf n_0$ denote the global normal vector of the IRS, where $\mathbf n_0$ is the local normal vector. The aperture projection factor associated with the BS--IRS--$X$ cascaded link is modeled as\cite{tang2022path}
	\begin{equation}
		F_X(\mathbf g)
		=
		\left[
		\frac{\mathbf n^T(\boldsymbol\gamma)(\mathbf p_B-\mathbf p_R)}
		{d_{B,R}}
		\right]_+
		\left[
		\frac{\mathbf n^T(\boldsymbol\gamma)(\mathbf p_X-\mathbf p_R)}
		{d_{R,X}}
		\right]_+,
		\label{eq:aperture}
	\end{equation}
	where $[x]_+=\max\{x,0\}$. For 	$ X\in\{U,E\}$, the equivalent BS--$X$ channel assisted by the UAV-mounted IRS is therefore
	\begin{equation}
		\mathbf h_X^H(\boldsymbol\Theta,\mathbf g)
		=
		\mathbf h_{B,X}^H
		+
		\sqrt{F_X(\mathbf g)}
		\mathbf h_{R,X}^H(\mathbf g)
		\boldsymbol\Theta
		\mathbf G_{B,R}(\mathbf g).
		\label{eq:equivalent_channel}
	\end{equation}
	
	
	
	\section{Problem Formulation}
In this paper, we consider the OFDM signal. In particular, the transmitted signal is given by $\mathbf{x} = \mathbf{U}^H\mathbf{s}$, where $\mathbf{U}$ is the normalized discrete Fourier transform (DFT) matrix of size $L$, and $\mathbf{s}\in\mathbb{C}^{L\times1}$ denotes the symbol vector. The $i$-th entry of $\mathbf{s}$ is randomly drawn from a complex constellation. In this paper, we assume that $\mathbb{E}(s_i) = 0$, $\mathbb{E}(|s_i|) = 1$. Let $\mathbf{f}\in\mathbb C^{N_t\times 1}$ denote the BS active beamforming vector. The received signal at the UE is
	\begin{equation}
		\mathbf{y}_U
		=
		\mathbf h_U^H(\boldsymbol\Theta,\mathbf g)\mathbf{f} \mathbf{x}^T+\mathbf{n}_U,
	\end{equation}
	where $\mathbf{n}_U\sim\CN(0,\sigma_U^2\mathbf{I})$. The corresponding communication signal-to-noise ratio (SNR) is
	\begin{equation}
		\gamma_U(\mathbf{f},\boldsymbol\Theta,\mathbf g)
		=
		\frac{
			|\mathbf h_U^H(\boldsymbol\Theta,\mathbf g)\mathbf{f}|^2
		}
		{\sigma_U^2}.
		\label{eq:gamma_u}
	\end{equation}
	
	The signal intercepted by Eve is
	\begin{equation}
		\mathbf{y}_E
		=
		\mathbf h_E^H(\boldsymbol\Theta,\mathbf g)\mathbf{f} \mathbf{x}^T+\mathbf{n}_E,
	\end{equation}
	where $\mathbf{n}_E\sim\CN(0,\sigma_E^2\mathbf{I})$. The eavesdropping SNR is
	\begin{equation}
		\gamma_E(\mathbf{f},\boldsymbol\Theta,\mathbf g)
		=
		\frac{
			|\mathbf h_E^H(\boldsymbol\Theta,\mathbf g)\mathbf{f}|^2
		}
		{\sigma_E^2}.
		\label{eq:gamma_e}
	\end{equation}
	
	
	The achievable secrecy rate is defined as
	\begin{equation}
		R_s
		=
		\left[
		\log_2(1+\gamma_U)
		-
		\log_2(1+\gamma_E)
		\right]^+.
	\end{equation}
	Since maximizing the nonnegative clipped secrecy rate is equivalent to maximizing the unclipped rate difference whenever a positive secrecy rate is achievable, we optimize
	\begin{equation}
		\bar R_s
		=
		\log_2(1+\gamma_U)
		-
		\log_2(1+\gamma_E).
	\end{equation}
	
	The considered secrecy-rate maximization problem is formulated as
	\begin{equation}
		\begin{split}
			\text{(P1)}\quad
			\max_{\mathbf{f},\mathbf v,\mathbf g}\quad
			&
			\log_2(1+\gamma_U)
			-
			\log_2(1+\gamma_E)\\
			\text{s.t.}\quad
			&
			\gamma_U(\mathbf{f},\boldsymbol\Theta,\mathbf g)\ge \Gamma_U,\\
			&
			\|\mathbf{f}\|_2^2\le P_{\max},\\
			&
			|v_n|=1,\quad n=1,\ldots,N,\\
			&
			\mathbf p_R\in\mathcal C,\quad
			\boldsymbol\gamma\in\mathcal G,\\
			&
			\mathbf n^T(\boldsymbol\gamma)(\mathbf p_X-\mathbf p_R)\ge 0,
			\; X\in\{B,U,E\}.
		\end{split}
		\label{eq:P1_visibility}
	\end{equation}
	Here, $\Gamma_U$ is the minimum communication QoS requirement, $P_{\max}$ is the maximum transmit power, $\mathcal C$ denotes the feasible UAV deployment region, and $\mathcal G$ denotes the allowable rotation range. Specifically, Constraint 1 guarantees the communication QoS requirement for the legitimate user (UE). Constraint 2 denotes the total transmit power limit at the BS, while constraint 3 represents the non-convex unit-modulus constraints for the IRS reflection. Constraint 4 defines the spatial and rotational boundaries for the 6-DoF movable IRS platform. Finally, constraint 5 ensures physical visibility, where the direction vector $\mathbf{p}_{X,R} = \mathbf{p}_X - \mathbf{p}_R$ from the IRS to node $X \in \{\text{BS, UE, Eve} \}$ must maintain an angle of less than 90 degrees relative to the panel normal \cite{liu2009secrecy}.
	
	Problem (P1) is difficult to solve globally because of the coupled active and passive beamforming variables, the unit-modulus constraints, and the highly nonlinear dependence of the channel on the IRS position and orientation. To obtain an efficient solution, we develop a three-block alternating optimization algorithm.
	
	\section{Proposed Alternating Optimization Algorithm}
	
	The proposed algorithm alternately updates the BS active beamforming vector $\mathbf{f}$, the IRS pose $\mathbf g$, and the IRS reflection vector $\mathbf v$. For notational simplicity, in each subproblem we omit variables that are fixed.
	
	\subsection{Active Beamforming Optimization via Linearized ADMM}
	
	For given $\mathbf{v}$ and $\mathbf{g}$, define 
    $\mathbf{A}_U = \frac{\mathbf{h}_U \mathbf{h}_U^H}{\sigma_U^2}$, and $ \mathbf{A}_E = \frac{\mathbf{h}_E \mathbf{h}_E^H}{\sigma_E^2}$. 
Then, we have 
    $\gamma_U=\mathbf{f}^H\mathbf A_U\mathbf{f},\quad
		\gamma_E=\mathbf{f}^H\mathbf A_E\mathbf{f}.$
	The active beamforming subproblem is
	\begin{equation}
		\begin{split}
			\max_{\mathbf{f}}\quad
			&
			\log_2(1+\mathbf{f}^H\mathbf A_U\mathbf{f})
			-
			\log_2(1+\mathbf{f}^H\mathbf A_E\mathbf{f})
			\\
			\text{s.t.}\quad
			&
			\mathbf{f}^H\mathbf A_U\mathbf{f}\ge \Gamma_U,\quad
			\|\mathbf{f}\|_2^2\le P_{\max}.
		\end{split}
	\end{equation}
		To facilitate a simple first-order update, we equivalently minimize the negativ secrecy objective and introduce an auxiliary variable $\mathbf z$ ith the consensus constraint $\mathbf{f}=\mathbf z$. The augmented  Lagrangian is
		\begin{equation}
		\begin{split}
			&\mathcal L_\rho(\mathbf{f},\mathbf z,\mathbf y)\\
			&=
			\; g_E(\mathbf{f})+g_U(\mathbf z)
			+\Real\{\mathbf y^H(\mathbf{f}-\mathbf z)\}
			+\frac{\rho}{2}\|\mathbf{f}-\mathbf z\|_2^2,
			\label{eq:aug_lagrangian}
		\end{split}
	\end{equation}
		where $g_E(\mathbf{f})=\log_2(1+\mathbf{f}^H\mathbf A_E\mathbf{f})$ and $g_U(\mathbf z)=-\log_2(1+\mathbf z^H\mathbf A_U\mathbf z)$,
		$\rho>0$ is the penalty parameter, and $\mathbf y$ is the dual variable.
		
		At the $(k+1)$-th inner iteration, the $\mathbf{f}$-update is obtained by linearizing $g_E(\mathbf{f})$ around $\mathbf{f}^{(k)}$:
		\begin{equation}
			\nabla_{\mathbf{f}^*}g_E(\mathbf{f}^{(k)})
			=
			\frac{1}{\ln 2}
			\frac{\mathbf A_E\mathbf{f}^{(k)}}
			{1+(\mathbf{f}^{(k)})^H\mathbf A_E\mathbf{f}^{(k)}}.
		\end{equation}
		%
     To simplify the mathematical representation, we define the intermediate gradient-step vector as
\begin{equation}
  \mathbf{b}^{(k)} = \mathbf{z}^{(k)} - \frac{1}{\rho} \left( \mathbf{y}^{(k)} + \nabla_{\mathbf{f}^*}g_E(\mathbf{f}^{(k)}) \right).
  \label{eq:b_definition}
\end{equation}
Then, the active beamforming vector is updated by
\begin{equation} 
  \mathbf{f}^{(k+1)} = \Pi_{\mathcal{W}} ( \mathbf{b}^{(k)} ),
  \label{eq:w_update} 
\end{equation} 
where $\Pi_{\mathcal{W}}(\cdot)$ denotes the projection operator onto the feasible set 
 \begin{equation} 
 \mathcal{W} = \{ \mathbf{f}: \mathbf{f}^H \mathbf{A}_U \mathbf{f} \ge \Gamma_U \}.
 \end{equation} 
The QoS-compliant projection $\Pi_{\mathcal{W}}(\mathbf{b})$ is mathematically defined as  \cite{xie2022perceptive}
\begin{equation} 
\Pi_{\mathcal{W}}(\mathbf{b}) = \begin{cases} 
\mathbf{b}, & \text{if } \mathbf{b}^H \mathbf{A}_U \mathbf{b} \ge \Gamma_U, \\ 
\mathbf{b}^{\perp} + \beta e^{j\phi} \bar{\mathbf{h}}_U, & \text{otherwise}, 
\end{cases} 
\label{eq:projection_cases} 
\end{equation} 
where $\bar{\mathbf{h}}_U = \mathbf{h}_U / \|\mathbf{h}_U\|_2$ is the maximum-ratio transmission direction, $\phi = \angle(\bar{\mathbf{h}}_U^H \mathbf{b})$, $\mathbf{b}^{\perp} = \mathbf{b} - (\bar{\mathbf{h}}_U^H \mathbf{b})\bar{\mathbf{h}}_U$, and $\beta = \sqrt{\frac{\Gamma_U \sigma_U^2}{\|\mathbf{h}_U\|_2^2}}$ is the minimum magnitude required to satisfy the QoS constraint $\Gamma_U$.

        For the $\mathbf z$-update, $g_U(\mathbf z)$ is linearized around $\mathbf z^{(k)}$. Its Wirtinger gradient  is \cite{wirtinger1927}
		\begin{equation}
			\nabla_{\mathbf z^*}g_U(\mathbf z^{(k)})
			=
			-
			\frac{1}{\ln 2}
			\frac{\mathbf A_U\mathbf z^{(k)}}
			{1+(\mathbf z^{(k)})^H\mathbf A_U\mathbf z^{(k)}}.
		\end{equation}
		Thus, the auxiliary variable is updated as
		\begin{equation}
			\mathbf z^{(k+1)}
			=
			\Pi_{\mathcal P}
			\left(
			\mathbf{f}^{(k+1)}
			+
			\frac{1}{\rho}
			\left(
			\mathbf y^{(k)}
			-
			\nabla_{\mathbf z^*}g_U(\mathbf z^{(k)})
			\right)
			\right),
			\label{eq:z_update}
		\end{equation}
		where $\mathcal P=\{\mathbf z:\|\mathbf z\|_2^2\le P_{\max}\}$. The dual variable is updated by
		\begin{equation}
			\mathbf y^{(k+1)}
			=
			\mathbf y^{(k)}
			+
			\rho(\mathbf{f}^{(k+1)}-\mathbf z^{(k+1)}).
			\label{eq:y_update}
		\end{equation}
		
		\subsection{Six-Dimensional Pose Optimization via Warm-Started PSO}
		
		For given $\mathbf{f}$ and $\mathbf v$, the IRS pose optimization subproblem is
		\begin{subequations}
			\begin{align}
				\max_{\mathbf g}\quad
				&
				R_s(\mathbf{f},\mathbf v,\mathbf g)
				\\
				\text{s.t.}\quad
				&
				\mathbf p_R\in\mathcal C,\quad
				\boldsymbol\gamma\in\mathcal G,
				\\
				&
				\mathbf n^T(\boldsymbol\gamma)(\mathbf p_X-\mathbf p_R)\ge 0,
				\quad X\in\{B,U,E\}.
			\end{align}
		\end{subequations}
		This subproblem is highly nonlinear because $\mathbf g$ affects the path loss, array responses, phase rotations, and aperture projection factors. To avoid poor local stationary points, we employ a warm-started PSO method.
		
		Each particle represents a candidate IRS pose:
		\begin{equation}
			\mathbf g_m
			=
			[p_{R,x},p_{R,y},p_{R,z},\gamma_x,\gamma_y,\gamma_z]^T,
			\quad m=1,\ldots,M.
		\end{equation}
		For the $m$-th particle, the penalized fitness function is defined as
		\begin{align}
			\mathcal F(\mathbf g_m)
			=
			&\;
			R_s(\mathbf{f},\mathbf v,\mathbf g_m)
			-
			\kappa_1
			\left[
			\Gamma_U-\gamma_U(\mathbf{f},\mathbf v,\mathbf g_m)
			\right]_+
			\nonumber\\
			&-
			\kappa_2
			\sum_{X\in\{B,U,E\}}
			\left[
			-\mathbf n^T(\boldsymbol\gamma)(\mathbf p_X-\mathbf p_R)
			\right]_+,
			\label{eq:fitness}
		\end{align}
		where $\kappa_1$ and $\kappa_2$ are penalty coefficients. The first penalty enforces the UE QoS requirement, while the second penalty enforces the physical visibility constraints.
		
		To improve search efficiency, the particle population is initialized by a hybrid strategy. In the first AO iteration, Latin hypercube sampling \cite{mckay2000comparison} is used to generate diverse initial poses within $\mathcal C\times\mathcal G$. In subsequent AO iterations, the best pose obtained in the previous AO round is used as the first particle, while the remaining particles are randomly sampled around it with bounded perturbations. This warm-start mechanism exploits the temporal correlation among AO iterations.
		
		Let $\boldsymbol\mu_m^{(t)}$ be the velocity of the $m$-th particle at PSO iteration $t$. The particle velocity and position are updated as
		\begin{align}
			\boldsymbol\mu_m^{(t+1)}
			=
			&\;
			\omega^{(t)}\boldsymbol\mu_m^{(t)}
			+
			c_1 r_1^{(t)}
			\left(
			\mathbf g_{m,\mathrm{pbest}}-\mathbf g_m^{(t)}
			\right)
			\nonumber\\
			&+
			c_2 r_2^{(t)}
			\left(
			\mathbf g_{\mathrm{gbest}}-\mathbf g_m^{(t)}
			\right),
			\label{eq:pso_velocity}\\
			\mathbf g_m^{(t+1)}
			=
			&\;
			\Pi_{\mathcal C\times\mathcal G}
			\left(
			\mathbf g_m^{(t)}
			+
			\boldsymbol\mu_m^{(t+1)}
			\right),
			\label{eq:pso_position}
		\end{align}
		where $c_1$ and $c_2$ are the cognitive and social learning factors, respectively, $r_1^{(t)}$ and $r_2^{(t)}$ are uniformly distributed random variables in $[0,1]$, and $\Pi_{\mathcal C\times\mathcal G}(\cdot)$ denotes element-wise projection onto the feasible pose region. The inertia weight $\omega^{(t)}$ is linearly adjusted as
		\begin{equation}
			\omega^{(t)}
			=
			\omega_{\max}
			-
			\frac{t}{T_{\rm PSO}}
			(\omega_{\max}-\omega_{\min}),
		\end{equation}
		where $T_{\rm PSO}$ is the maximum number of PSO iterations.
		
		\subsection{Passive Beamforming (PBF) Optimization on the Complex-Circle Manifold}
		
		For given $\mathbf{f}$ and $\mathbf g$, the passive reflection design is
		\begin{subequations}
			\begin{align}
				\max_{\mathbf v}\quad
				&
				R_s(\mathbf{f},\mathbf v,\mathbf g)
				\\
				\text{s.t.}\quad
				&
				|v_n|=1,\quad n=1,\ldots,N.
			\end{align}
		\end{subequations}
		The feasible set is the product of $N$ complex circles:
		\begin{equation}
			\mathcal M
			=
			\left\{
			\mathbf v\in\mathbb C^N:
			|v_n|=1,\; n=1,\ldots,N
			\right\}.
		\end{equation}
		Thus, the passive beamforming subproblem can be efficiently solved by Riemannian gradient descent.
		
		For $X\in\{U,E\}$, the equivalent channel in \eqref{eq:equivalent_channel} can be rewritten as an affine function of $\mathbf v$:
		\begin{equation}
			\mathbf h_X^H
			=
			\mathbf h_{B,X}^H
			+
			\mathbf v^T\mathbf D_X,
			\label{eq:affine_v}
		\end{equation}
		where
		\begin{equation}
			\mathbf D_X
			=
			\sqrt{F_X(\mathbf g)}
			\diag(\mathbf h_{R,X}^H(\mathbf g))
			\mathbf G_{B,R}(\mathbf g).
		\end{equation}
		Define
		\begin{equation}
			q_X(\mathbf v)
			=
			\mathbf h_X^H\mathbf{f}
			=
			\mathbf h_{B,X}^H\mathbf{f}
			+
			\mathbf v^T\mathbf D_X\mathbf{f}.
		\end{equation}
		Then, we have
		\begin{equation}
			\gamma_X(\mathbf v)=\frac{|q_X(\mathbf v)|^2}{\sigma_X^2},
			\quad X\in\{U,E\}.
		\end{equation}
		The Euclidean gradient of the negative secrecy objective
		\begin{equation}
			f_v(\mathbf v)
			=
			\log_2(1+\gamma_E(\mathbf v))
			-
			\log_2(1+\gamma_U(\mathbf v))
		\end{equation}
		with respect to $\mathbf v^*$ is
		\begin{align}
			\nabla_{\mathbf v^*} f_v
			=
			&
			\frac{1}{\ln 2}
			\frac{1}{1+\gamma_E}
			\nabla_{\mathbf v^*}\gamma_E
			-
			\frac{1}{\ln 2}
			\frac{1}{1+\gamma_U}
			\nabla_{\mathbf v^*}\gamma_U,
			\label{eq:euc_grad_v}
		\end{align}
		where
		\begin{equation}
			\nabla_{\mathbf v^*}\gamma_X
			=
			\frac{1}{\sigma_X^2}
			\left(\mathbf D_X\mathbf{f}\right)^*
			q_X(\mathbf v),
			\quad X\in\{U,E\}.
		\end{equation}
		
		The tangent space of $\mathcal M$ at $\mathbf v$ is
		\begin{equation}
			T_{\mathbf v}\mathcal M
			=
			\left\{
			\boldsymbol\xi\in\mathbb C^N:
			\Real\{\boldsymbol\xi\odot \mathbf v^*\}=\mathbf 0
			\right\}.
		\end{equation}
		The Riemannian gradient is obtained by projecting the Euclidean gradient onto the tangent space:
		\begin{equation}
			\operatorname{grad} f_v
			=
			\nabla_{\mathbf v^*} f_v
			-
			\Real\left\{
			\nabla_{\mathbf v^*} f_v\odot \mathbf v^*
			\right\}
			\odot \mathbf v.
			\label{eq:riem_grad}
		\end{equation}
		Given a step size $\alpha_i$, the retraction operation\cite{absil2008optimization} is
		\begin{equation}
			\begin{split}
			\mathbf v^{(i+1)}
			&=
			\operatorname{Retr}_{\mathbf v^{(i)}}
			\left(
			-\alpha_i\operatorname{grad}f_v(\mathbf v^{(i)})
			\right)\\
			&=
			e^{\jmath\arg\left(
				\mathbf v^{(i)}
				-
				\alpha_i\operatorname{grad}f_v(\mathbf v^{(i)})
				\right)}.
			\label{eq:retraction}
			\end{split}
		\end{equation}
		The step size is selected by backtracking line search to guarantee monotonic improvement of the secrecy objective.
		
		
        \emph{Remark:} Algorithm 1 summarizes the AO procedure, which guarantees convergence since the objective value is monotonically non-decreasing and upper-bounded. Specifically, updating active beamformer $\mathbf{f}$ (with auxiliary $\mathbf{z}$) requires $\mathcal{O}(I_{\text{ADMM}} N_t^2)$ operations; searching 6D pose $\mathbf{g}$ requires $\mathcal{O}(T_{\text{PSO}} M N N_t)$ operations; and updating passive vector $\mathbf{v}$ requires $\mathcal{O}(I_{\text{RGD}} N N_t)$ operations. Thus, the total per-iteration complexity is $\mathcal{O}(I_{\text{ADMM}} N_t^2 + T_{\text{PSO}} M N N_t + I_{\text{RGD}} N N_t)$, where $I_{\text{ADMM}}$, $N_t$, $T_{\text{PSO}}$, $M$, $N$, and $I_{\text{RGD}}$ denote the ADMM inner iterations, BS antennas, maximum PSO iterations, particle size, IRS elements, and backtracking steps, respectively. This polynomial complexity underscores the practical feasibility of the proposed algorithm for real-time deployment.
		
		\begin{algorithm}[t]
			\caption{Proposed AO Algorithm for Secure 6D Movable IRS-Assisted ISAC}
			\label{alg:overall}
			\begin{algorithmic}[1]
				\STATE \textbf{Input:} Locations $\mathbf p_B,\mathbf p_U,\mathbf p_E$, feasible region $\mathcal C$, rotation set $\mathcal G$, power budget $P_{\max}$, QoS threshold $\Gamma_U$.
				\STATE Initialize $\mathbf{f}^{(0)}$, $\mathbf v^{(0)}$, and $\mathbf g^{(0)}$.
				\STATE Set AO iteration index $r=0$.
				\REPEAT
				\STATE Given $\mathbf v^{(r)}$ and $\mathbf g^{(r)}$, update $\mathbf{f}^{(r+1)}$ by the linearized ADMM iterations in \eqref{eq:w_update}--\eqref{eq:y_update}.
				\STATE Given $\mathbf{f}^{(r+1)}$ and $\mathbf v^{(r)}$, update $\mathbf g^{(r+1)}$ by the warm-started PSO in \eqref{eq:pso_velocity}--\eqref{eq:pso_position}.
				\STATE Given $\mathbf{f}^{(r+1)}$ and $\mathbf g^{(r+1)}$, update $\mathbf v^{(r+1)}$ by Riemannian gradient descent in \eqref{eq:riem_grad}--\eqref{eq:retraction}.
				\STATE Compute $R_s^{(r+1)}$.
				\STATE $r\leftarrow r+1$.
				\UNTIL{$|R_s^{(r)}-R_s^{(r-1)}|\le \epsilon$ or $r=R_{\max}$.}
				\STATE \textbf{Output:} $\mathbf{f}^\star$, $\mathbf v^\star$, $\mathbf g^\star$.
			\end{algorithmic}
		\end{algorithm}

		\section{Simulation Results}
		
		In this section, numerical simulations are presented to evaluate the proposed UAV-mounted six-dimensional movable IRS-assisted secure ISAC design. Unless otherwise specified, the BS is equipped with $N_t=32$ transmit antennas and $N_r=32$ receive antennas. The IRS is configured as a UPA with $N=N_xN_z$ reflecting elements. The carrier frequency is $f_c=3.6$ GHz. The reference path loss is set to $\beta_0=-30$ dB. The noise power spectral density is $-174$ dBm/Hz, and the system bandwidth is $1$ MHz. The maximum transmit power is $P_{\max}=30$ dBm.
		The BS is located at $\mathbf{p}_B=[0,0,0]^T$ m. The legitimate user and the potential eavesdropper/sensing target are located at $\mathbf{p}_U=[280,0,0]^T$ m and $\mathbf{p}_E=[0,20,0]^T$ m, respectively. The UAV-mounted IRS flies at a fixed altitude $H=150$ m, i.e., $p_{R,z}=H$. The path-loss exponents of the communication and eavesdropping links are set to $3$ and $2.2$, respectively. The PSO parameters are set as $c_1=0.9$, $c_2=0.1$, $\omega_{\min}=0.1$, and $\omega_{\max}=1.1$.
		
		To demonstrate the impact of the six-dimensional mobility, we compare the following schemes:
		\begin{itemize}
			\item \textbf{Proposed 6D+PBF, large region:} The IRS six-dimensional pose and passive beamforming are jointly optimized in a large deployment region $\mathcal R_2$, where $p_{R,x}\in[0,100]$ m, and $p_{R,y}\in[0,100]$ m.
			\item \textbf{Proposed 6D+PBF, small region:} The IRS six-dimensional pose and passive beamforming are jointly optimized in a smaller region $\mathcal R_1$, where $p_{R,x}\in[50,100]\;{\rm m}$, and $p_{R,y}\in[50,100]\;{\rm m}$.
			\item \textbf{3D orientation+PBF:} The IRS location is fixed at $\mathbf{p}_R = [100, 100, 150]^T$~m, while only its three-dimensional orientation and passive reflection coefficients are optimized.
			\item \textbf{Fixed IRS baseline:} The IRS location is fixed at $\mathbf{p}_R = [100, 100, 150]^T$~m, and its orientation is fixed as $\boldsymbol{\gamma} = \left[ \frac{5}{3}\pi, \frac{7}{5}\pi, \frac{1}{7}\pi \right]^T$, and only the passive reflection coefficients are optimized.
		\end{itemize}

		\begin{figure}[t]
			\centering
			\includegraphics[width=0.92\linewidth]{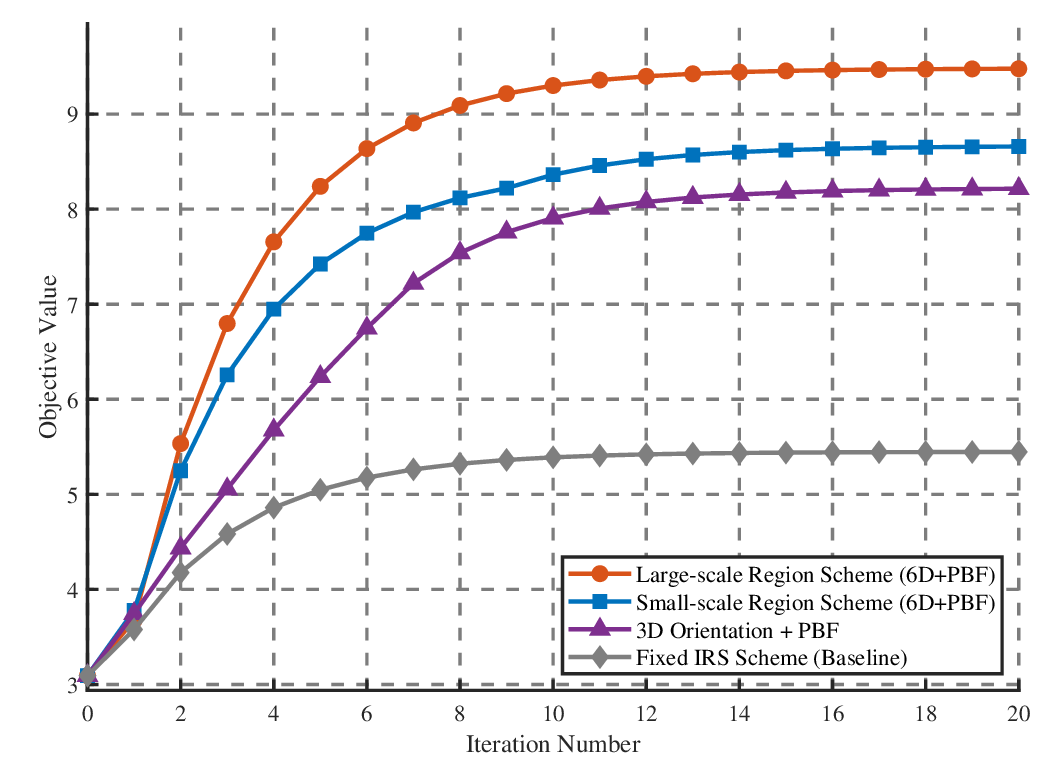}
			\caption{Convergence behavior of the proposed algorithm under different 3D spatial constraints ($N = 16$).}
			\label{fig:region_comparison}
		\end{figure}
		

       Fig.~\ref{fig:region_comparison} compares the convergence performance under different spatial mobility constraints. For all schemes, the objective value increases rapidly during the first few AO iterations and then stabilizes. This confirms that the proposed block-wise design effectively coordinates active beamforming, 6D pose control, and passive reflection optimization. The early rapid improvement is driven by the macroscopic pose adjustment of the UAV-mounted IRS which substantially changes the channel geometry, while the later fine-grained gains stem from passive phase alignment on the IRS. Furthermore, the proposed 6D+PBF schemes outperform the orientation-only and fixed-IRS baselines. Specifically, the large-region scheme achieves a higher secrecy rate than the small-region counterpart, as a larger feasible region provides more spatial DoFs to balance communication enhancement and eavesdropping suppression. This reveals that the translation DoFs of the UAV-mounted IRS are essential for secure ISAC, especially when the UE and Eve have close angular directions relative to the BS.

		\begin{figure}[t]
			\centering
			\includegraphics[width=0.92\linewidth]{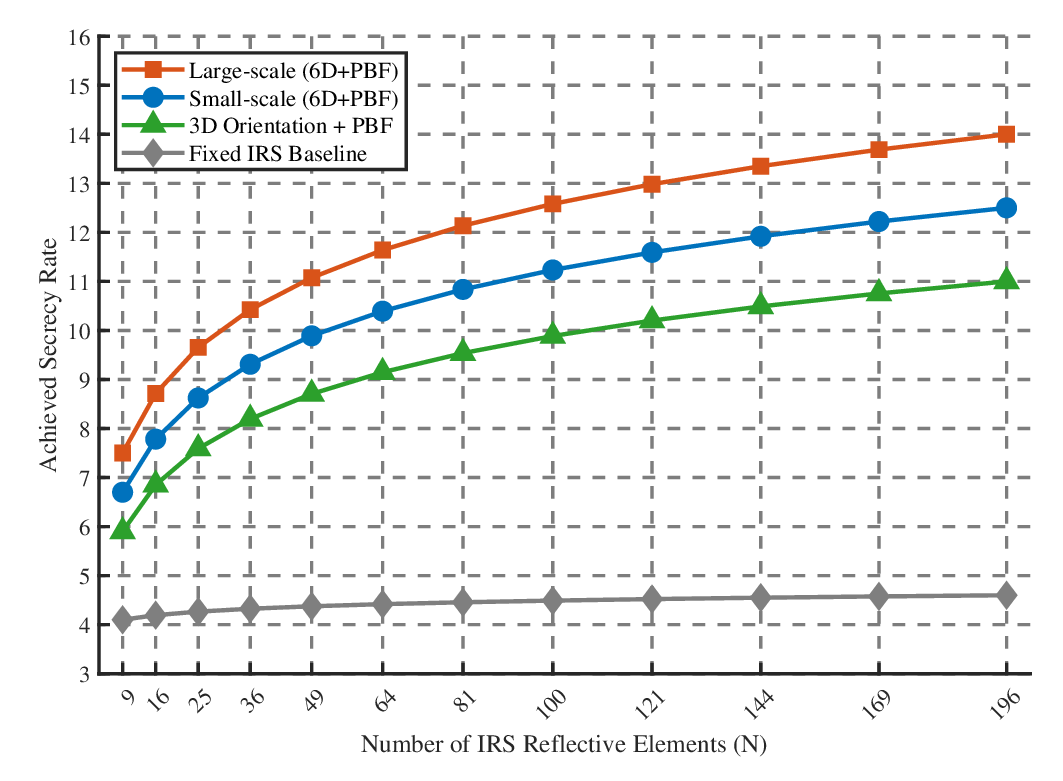}
			\caption{Achievable secrecy rate versus the number of IRS reflecting elements, $N$, for different configurations.}
			\label{fig:N_comparison}
		\end{figure}
		
		Fig.~\ref{fig:N_comparison} illustrates the impact of the number of IRS reflective elements, $N$, on the achievable secrecy rate. 
        As $N$ increases, the secrecy rate exhibits a clear upward trend, credited to the enhanced passive aperture gain and sharper beam directivity.
        However, as $N$ grows beyond a certain threshold, the performance gain gradually saturates. This demonstrates that in line-of-sight (LoS)-dominated propagation environments, the system performance faces an inherent physical bottleneck, and solely expanding the IRS size cannot yield unlimited security enhancements. 
        Additionally, the performance relations among different configurations remain consistent with our previous observations. 
        Specifically, the 6D+PBF scheme excels over both baselines, while the larger region surpasses its smaller counterpart, reinforcing the efficacy of the 6D+PBF design and relaxed spatial constraints in boosting secure transmission.

       Overall, the simulation results confirm that 6D pose optimization provides a key spatial-domain mechanism for enhancing physical-layer security, particularly in the angular-overlap regime where the UE and Eve are closely aligned from the perspective of the BS.  In this regime, their highly correlated direct channels cause unavoidable leakage and limit the effectiveness of conventional fixed IRSs. By contrast, the proposed 6D movable IRS exploits 3D translation to reshape the path loss and angular geometry, thereby improving link separability, while using 3D rotation to adjust the panel normal and local propagation angles. These two mechanisms significantly amplify the disparity between their effective aperture projection factors ($F_{\text{U}}(\mathbf{g})$ and $F_{\text{E}}(\mathbf{g})$), creating a favorable power-domain asymmetry that enhances the legitimate cascaded link.

		\section{Conclusion}
		

       This paper investigated a UAV-mounted 6D movable IRS-assisted secure ISAC system. By jointly optimizing the active beamformer, 6D IRS pose, and passive reflection vector, we maximized the secrecy rate under QoS and visibility constraints. An efficient alternating optimization framework was proposed based on linearized ADMM, warm-started PSO, and Riemannian gradient descent. Simulation results demonstrated that the proposed scheme achieves significant secrecy gains over static baselines.Future work may extend this framework to multi-user secure ISAC, robust design under imperfect channel state information, and dynamic UAV trajectory optimization.
		
		\bibliographystyle{IEEEtran}
    \bibliography{ref}

@article{wang2025uav,
  author  = {Wang, P. and Xue, Y. and Mei, W. and Fang, J. and Zhang, R.},
  title   = {{UAV}-enabled passive {6D} movable antenna for {ISAC}: Joint location, orientation, and reflection optimization},
  journal = {IEEE Wireless Commun. Lett.},
  note    = {Early access},
  year    = {2025}
}

@article{liu2009secrecy,
  author  = {Liu, T. and Shamai, S.},
  title   = {A note on the secrecy capacity of the multiple-antenna wiretap channel},
  journal = {IEEE Trans. Inf. Theory},
  volume  = {55},
  number  = {6},
  pages   = {2547--2553},
  month   = jun,
  year    = {2009}
}

@article{boyd2011distributed,
  author  = {Boyd, S. and Parikh, N. and Chu, E. and Peleato, B. and Eckstein, J.},
  title   = {Distributed optimization and statistical learning via the alternating direction method of multipliers},
  journal = {Found. Trends Mach. Learn.},
  volume  = {3},
  number  = {1},
  pages   = {1--122},
  year    = {2011}
}

@article{tang2022path,
 author={Tang, Wankai and Chen, Xiangyu and Chen, Ming Zheng and Dai, Jun Yan and Han, Yu and Di Renzo, Marco and Jin, Shi and Cheng, Qiang and Cui, Tie Jun},
  title={Path loss modeling and measurements for reconfigurable intelligent surfaces in the millimeter-wave frequency band},
  journal={IEEE Transactions on Communications},
  volume={70},
  number={9},
  pages={6259--6276},
  year={2022},
  publisher={IEEE}
}

@article{wirtinger1927,
  author  = {Wirtinger, W.},
  title   = {Zur formalen Theorie der Funktionen von mehr komplexen {Ver{\"a}nderlichen}},
  journal = {Math. Ann.},
  volume  = {97},
  number  = {1},
  pages   = {357--375},
  month   = dec,
  year    = {1927}
}

@inproceedings{kennedy1995particle,
  author    = {Kennedy, J. and Eberhart, R.},
  title     = {Particle swarm optimization},
  booktitle = {Proc. IEEE Int. Conf. Neural Networks},
  address   = {Perth, WA, Australia},
  pages     = {1942--1948},
  month     = nov,
  year      = {1995}
}

@article{mckay2000comparison,
  author  = {McKay, M. D. and Beckman, R. J. and Conover, W. J.},
  title   = {A comparison of three methods for selecting values of input variables in the analysis of output from a computer code},
  journal = {Technometrics},
  volume  = {42},
  number  = {1},
  pages   = {55--61},
  month   = feb,
  year    = {2000}
}

@book{absil2008optimization,
  author    = {Absil, P.-A. and Mahony, R. and Sepulchre, R.},
  title     = {Optimization Algorithms on Matrix Manifolds},
  publisher = {Princeton Univ. Press},
  address   = {Princeton, NJ, USA},
  year      = {2008}
}

@article{xie2026secure,
  title={Secure Communication in {MIMOME} Movable-Antenna Systems with Statistical Eavesdropper CSI},
  author={Xie, Lei and Wang, Peilan and Shen, Guanxiong and Li, Guyue and Mei, Weidong and Chen, Liquan},
  journal={arXiv preprint arXiv:2601.14755},
  year={2026}
}

@article{xie2022perceptive,
  title={Perceptive mobile network with distributed target monitoring terminals: Leaking communication energy for sensing},
  author={Xie, Lei and Wang, Peilan and Song, SH and Letaief, Khaled B},
  journal={IEEE Trans. Wireless Commun.},
  volume={21},
  number={12},
  pages={10193--10207},
  year={2022},
  publisher={IEEE}
}

@article{liu2020jointradarcomm,
  author  = {F. Liu and C. Masouros and A. P. Petropulu and H. Griffiths and L. Hanzo},
  title   = {Joint Radar and Communication Design: Applications, State-of-the-Art, and the Road Ahead},
  journal = {IEEE Trans. Commun.},
  volume  = {68},
  number  = {6},
  pages   = {3834--3862},
  year    = {2020},
  month   = jun
}

@article{xie2023collaborativesensing,
  author  = {L. Xie and S. Song and Y. C. Eldar and K. B. Letaief},
  title   = {Collaborative Sensing in Perceptive Mobile Networks: Opportunities and Challenges},
  journal = {IEEE Wireless Commun.},
  volume  = {30},
  number  = {1},
  pages   = {16--23},
  year    = {2023}
}

@article{wu2025irsisacsurvey,
  title={Intelligent reflecting surfaces for integrated sensing and communications: A survey},
  author={Wu, Qingqing and Peng, Qiaoyan and Zhang, Ziheng and Shao, Xiaodan and Liu, Yang and Jiang, Yifan and Zhao, Yapeng and Zhu, Yanze and Chen, Yilong and Ren, Zixiang and others},
  journal={arXiv preprint arXiv:2511.10990},
  year={2025}
}

@article{wu2020towardsris,
  author  = {Q. Wu and R. Zhang},
  title   = {Towards Smart and Reconfigurable Environment: Intelligent Reflecting Surface Aided Wireless Network},
  journal = {IEEE Commun. Mag.},
  volume  = {58},
  number  = {1},
  pages   = {106--112},
  year    = {2020},
  month   = jan
}

@article{pang2022irssecureuav,
  author  = {X. Pang and N. Zhao and J. Tang and C. Wu and D. Niyato and K.-K. Wong},
  title   = {{IRS-Assisted} Secure {UAV} Transmission via Joint Trajectory and Beamforming Design},
  journal = {IEEE Trans. Commun.},
  volume  = {70},
  number  = {2},
  pages   = {1140--1152},
  year    = {2022},
  month   = feb
}

@article{zhou2025rotatableirs6dma,
  title={Rotatable {IRS-Assisted} {6DMA} Communications: A Two-timescale Design},
  author={Zhou, Chao and You, Changsheng and Zhou, Cong and Yao, Liujia and Yuan, Weijie and Zheng, Beixiong and Wu, Nan},
  journal={arXiv preprint arXiv:2512.15092},
  year={2025}
}

@article{wu2025mairs,
  title={Integrating movable antennas and intelligent reflecting surfaces {(MA-IRS)}: Fundamentals, practical solutions, and {ISAC}},
  author={Wu, Qingqing and Zheng, Ziyuan and Gao, Ying and Mei, Weidong and Wei, Xin and Chen, Wen and Ning, Boyu},
  journal={IEEE Wireless Commun.},
  year={2025},
  publisher={IEEE}
}
		
	\end{document}